\documentclass{jfm}

\usepackage[dvips]{graphicx}% Include figure files
\usepackage{natbib}
\usepackage{bm}% bold mathÄ
\usepackage[normalem]{ulem}
\usepackage[usenames]{color}
\usepackage{psfrag}
\usepackage{textcomp}
\usepackage{amsmath}

\def\gtwid{\mathrel{\raise.3ex\hbox{$>$\kern-.75em\lower1ex\hbox{$\sim$}}}}
\def\alt{\mathrel{\raise.3ex\hbox{$<$\kern-.75em\lower1ex\hbox{$\sim$}}}}
\def\agt{\mathrel{\raise.3ex\hbox{$>$\kern-.75em\lower1ex\hbox{$\sim$}}}}

\newcommand{\be}{\begin{equation}}
\newcommand{\ee}{\end{equation}}

\begin{document}

\title[Azimuthal diffusion of the large-scale-circulation plane]{Azimuthal diffusion of the large-scale-circulation plane, and absence of significant non-Boussinesq effects, in turbulent convection near the ultimate-state transition}

\author{Xiaozhou He$^{1,2,6}$, Eberhard Bodenschatz$^{1,3,4,6}$, and Guenter Ahlers$^{1,5,6}$}

\affiliation{
$^{1}$Max Planck Institute for Dynamics and Self Organization (MPIDS), D-37073 G\"ottingen, Germany
\\
$^{2}$Institute for Turbulence-Noise-Vibration Interaction and Control, Shenzhen Graduate School, Harbin Institute of Technology, Shenzhen, China
\\
$^{3}$Institute for Nonlinear Dynamics, University of G\"ottingen, D-37073 G\"ottingen, Germany
\\
$^{4}$Laboratory of Atomic and Solid-State Physics and Sibley School of Mechanical and Aerospace Engineering, Cornell University, Ithaca, New York 14853
\\
$^5$Department of Physics, University of California, Santa Barbara, California 93106
\\
$^{6}$International Collaboration for Turbulence Research
}

\date{\today}
\maketitle

\begin{abstract}

We present measurements of the orientation $\theta_0$ and temperature amplitude $\delta$ of the large-scale circulation in a cylindrical sample of turbulent Rayleigh-Benard convection (RBC) with aspect ratio $\Gamma \equiv D/L = 1.00$ ($D$ and $L$ are the diameter and height respectively) and for the Prandtl number $Pr \simeq 0.8$. Results for $\theta_0$  revealed a preferred orientation with upflow in the West, consistent with a broken azimuthal invariance due to Earth's Coriolis force [see \cite{BA06b}]. They yielded the azimuthal diffusivity $D_\theta$ and a corresponding Reynolds number $Re_{\theta}$ for Rayleigh numbers over the range $2\times 10^{12} \alt Ra \alt 1.5\times 10^{14}$. In the classical state ($Ra \alt 2\times 10^{13}$) the results were consistent with the measurements by \cite{BA06a} for $Ra \alt 10^{11}$ and $Pr = 4.38$ which gave $Re_{\theta} \propto Ra^{0.28}$, and with the Prandtl-number dependence $Re_{\theta} \propto Pr^{-1.2}$ as found previously also for the velocity-fluctuation Reynolds number $Re_V$ \cite[]{HGBA15b}. At larger $Ra$ the data for $Re_{\theta}(Ra)$ revealed a transition to a new state, known as the ``ultimate" state, which was first seen in the Nusselt number $Nu(Ra)$ and in $Re_V(Ra)$ at $Ra^*_1 \simeq 2\times 10^{13}$ and $Ra^*_2 \simeq 8\times 10^{13}$. In the ultimate state we found $Re_{\theta} \propto Ra^{0.40\pm 0.03}$.

Recently \cite{SU15} claimed that non-Oberbeck-Boussinesq effects on the Nusselt and Reynolds numbers of turbulent RBC may have been interpreted erroneously as a transition to a new state. We demonstrate that their reasoning is incorrect and that the transition observed in the G\"ottingen experiments and discussed in the present paper is indeed to a  new state of RBC referred to as ``ultimate".   

\end{abstract}

\section{Introduction}

Turbulent convection in a fluid contained between two parallel horizontal plates and heated from below [Rayleigh-B\'enard convection or RBC; for reviews see \cite{Ka01,Ah09, AGL09, LX10,CS12}] contains two thin laminar boundary layers (BLs), one below the top and the other above the bottom plate, when the Rayleigh number $Ra$ is not too large. When the fluid properties (except for the density where its temperature dependence drives the flow) are temperature independent [Oberbeck-Boussinesq (OB) conditions \cite[]{Ob79,Bo03}], then both BLs are of equal thickness and each sustains an equal temperature difference nearly equal to $\Delta T/2$ where $\Delta T \equiv T_b - T_t$. Here $T_b$ and $T_t$ are the temperatures at the bottom and top plate respectively. The sample center temperature then is $T_c = T_m$  where $T_m \equiv (T_b + T_t)/2$. 
The fluid not too close to the BLs, known as the ``bulk", is strongly turbulent but nearly isothermal in the time average [see, however, \cite{ABFGHLSV12,ABH14,WA14}], with a vigorously fluctuating temperature \cite[]{HGBA13}. The BLs emit thermal plumes at irregular time intervals which rise or fall through the bulk and, by virtue of their buoyancy, drive and in turn are carried by a large-scale circulation (LSC). For cylindrical samples with height $L$ about equal to the diameter  $D$ (aspect ratio $\Gamma \equiv D/L \simeq 1$) the LSC consists of a single convection roll with rising fluid near the wall at an azimuthal location $\theta_0$ and falling  fluid also near the wall but close to $\theta_0 + \pi$. 

The state of the system depends on 
$Ra \equiv g\alpha\Delta T L^3/(\kappa \nu)$
and the Prandtl number
$Pr \equiv \nu/\kappa$.
Here $g$, $\alpha$, $\kappa$ and $\nu$ denote the gravitational acceleration, the isobaric thermal expansion coefficient, the thermal diffusivity, and the kinematic viscosity respectively. 
The system described in the previous paragraph is referred to as ``classical" RBC. When $Ra$ reaches a certain value $Ra^*$, it is expected theoretically \cite[]{Kr62,Sp71,GL11} that the laminar BLs of the classical state will also become turbulent due to the shear applied by the LSC, as well as by fluctuations on somewhat smaller scales. Above $Ra^*$ the system is said to be in the ``ultimate" state as this state is expected to persist up to arbitrarily high $Ra$ \cite[]{CCCHCC97}. A theoretical estimate gave $Ra^* = {\cal O}(10^{14})$ for $Pr$ near one \cite[]{GL02}. Several experimental papers  reported transitions found from measurements of the  heat transport, expressed as the Nusselt number 
$Nu= \lambda_{eff}/\lambda$
where $\lambda$ is the thermal conductivity of the quiescent fluid and 
$\lambda_{eff}=Q L/(A \Delta T)$. 
Here $Q$ is the heat flux and $A$ the cross sectional area of the cell.
The results of \cite{CCCHCC97}, \cite{CCCCH01}, and \cite{RGKS10} (the ``Grenoble" data, obtained with fluid helium at a temperature of about 5 Kelvin and a pressure of about 2 bars) found transitions in the $Ra$ range from $10^{11}$ to $10^{12}$, which in our view (but not that of the authors) is too low to correspond to a BL shear instability although the data clearly show a sharp and continuous transition which, we believe, is of unknown origin (see \cite{AHFB12}). Measurements of $Nu$ and of the Reynolds number $Re_V = V L / \nu$ ($V$ is the root-mean-square fluctuation velocity) made with compressed sulfur hexafluoride (SF$_6$) at ambient temperatures and pressures up to 19 bars also found a transition [\cite{HFNBA12,AHFB12,HFBA12,HGBA15b}, the ``G\"ottingen" data], but at $Ra^* \simeq 10^{14}$ in agreement with the theoretical estimate of \cite{GL02}.   

Here we present a study of the azimuthal diffusivity $D_{\theta}$ of the LSC ($\S$ \ref{sec:LSC}). 
The LSC is a stochastically driven system, with the driving due to the small-scale fluctuations \cite[]{BA07a,BA08a}. In view of the rotationally invariant geometry of the cylindrical sample $\theta_0$ ideally should diffuse azimuthally in the presence of the stochastic driving and have a uniform probability distribution $p(\theta_0) = 1/(2\pi)$.
However, in real experiments $p(\theta_0)$ always is found to have a maximum even when extraordinary care is used in preparing the sample cell, indicating that the rotational symmetry is broken by some external force or remaining internal imperfection. 
Nearly a decade ago it was shown that the symmetry breaking by Earth's Coriolis force is sufficient to yield the $p(\theta_0)$ observed in an experiment \cite[]{BA06b}. A simple Navier-Stokes based model for the azimuthal potential due to the Coriolis force revealed that $\theta_0$ should be very close to West. Solving a Fokker-Planck equation with this model potential and a noise intensity derived from the measured $D_{\theta}$  yielded a $p(\theta_0)$ which agreed quantitatively with experiment. For that case the Prandtl number was 4.38 and $\Gamma = 1.00$. Two samples were investigated, one with $L = 24.8$ and the other with $L \simeq 50$ cm. The Rayleigh number was in the range $3\times 10^8 \alt Ra \alt 10^{11}$.

The present study is for a physically larger sample,  with $L = D = 112$ cm, and with the smaller $Pr \simeq 0.8$. Rayleigh numbers ranged from $2\times 10^{12}$ to $1.5\times 10^{14}$ and thus were much larger than in the previous work.  Again we found that the preferred $\theta_0$ was close to  West, consistent with Earth's Coriolis force. We also measured $D_\theta$ and determined the corresponding Reynolds number $Re_{\theta} = L \sqrt{D_\theta/\nu}$. Comparison with an extrapolation of the earlier data indicates consistency with $Re_{\theta} \propto Pr^{1.2}$, a Prandtl dependence found before for $Re_V$ \cite[]{HGBA15b}. 
Quite remarkably, the diffusivity measurements confirmed that the ultimate-state transition occurs over a {\it range} of $Ra$, from 
 $Ra^*_1 \simeq 2\times 10^{13}$ to $Ra^*_2 \simeq 8\times 10^{13}$, as was found earlier from measurements of $Nu(Ra)$ and $Re_V(Ra)$. 

We also present new measurements of $Nu$ which are for the same sample as that used for the study of the azimuthal diffusion. They, as well as $Re_V$ measurements reported before \cite[]{HGBA15b} for a different sample with $\Gamma = 1.00$, show the ultimate-state transition and all three physical properties yield consistent results for $Ra^*_1$ and $Re^*_2$ as shown below in Fig.~\ref{fig:Re_th_Pr}.   

Recently it was claimed by \cite{SU15} that the temperature dependence of fluid properties (non-OB conditions) may be responsible for the transitions seen in the Grenoble and the G\"ottingen data and that hypothetical corresponding data for an OB system may not show any transitions. In $\S$ \ref{sec:NOB} we show two reasons why this claim is flawed.

\section{Apparatus and procedure}
\label{sec:App}

The apparatus was described elsewhere \cite[]{AFB09,AHFB12,ABH14}. 
It consisted of a sample cell located inside a vessel known as the ``Uboot of G\"ottingen" which was pressurized with sulfur hexafluoride (SF$_6$) at an average temperature $T_{m} = 21.5$\textcelsius\ and at pressures that ranged from 2.0 to 
17.7 bars, resulting in the relatively narrow Prandtl-number range from 0.78 to 0.86 over the $Ra$ range from $2\times 10^{12}$ to $1.5\times 10^{14}$ (see Fig. 1b of \cite{ABH14}).  The Uboot had a volume of about 25 m$^3$, and about 2000 kg of SF$_6$ were required to fill it to the maximum pressure. The sample cell had  aluminium plates at the top and bottom that were separated by a distance $L = 112$ cm and had an aspect ratio  $\Gamma = 1.00$. The plates were leveled relative to gravity to $\pm 10^{-4}$ rad. 

Three sets of eight thermistors, one each at the heights $z/L = 0.25$, 0.50, and 0.75, were imbedded in the sidewall \cite[]{BNA05}. Since the  LSC carries warm fluid upward at one side of the cell and cold fluid downward at the opposite, one can detect the azimuthal orientation of the LSC by fitting 
\begin{equation}
T_f=T_{w,k}+\delta_k \cos \left( \textrm{i}\pi/4-\theta_{k,0} \right)
\label{eq:cos}
\end{equation} 
to all eight thermistors at one specific height and time. The index ``i" stands for the azimuthal location of the thermistors and takes values $i=0\dots 7$, with the location of $i = 0$ located at a point directly East of the sample center and $i$ increasing in the counter-clockwise direction when viewed from above. The index $k$ denotes the vertical location of the thermistor and will in the following have letters $b$ ($z=L/4$), $m$ ($z=L/2$) and $t$ ($z=3L/4$). The orientation of the up-flow is given by the phase $\theta_{k,0}$.

From the eight temperatures at a given $k$ we computed  the coefficients of the lowest four Fourier modes \cite[]{SCL11,WA11c}.  This yielded all eight Fourier coefficients $A_{k,j}$ (the cosine coefficients) and $B_{k,j}$ (the sine coefficients), $j = 1, ... 4$ and the corresponding ``energy" $E_{k,j}(t) = A_{k,j}^2 + B_{k,j}^2$. The time averages $\langle E_{k,j}(t)\rangle$ are summed to compute the total energy $E_{k,tot}$.

\section{Results}
\label{sec:app+proc}

\subsection{Orientation and energy of the large-scale circulation}
\label{sec:LSC}

\begin{figure}
\centerline{\includegraphics[width=0.85 \textwidth]{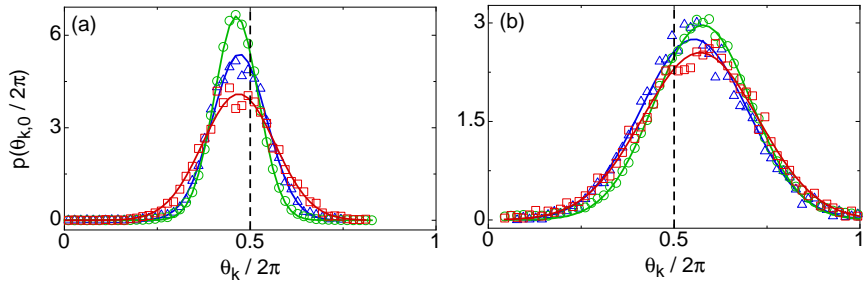}}
\caption{The probability distribution $p(\theta_{k,0} / 2 \pi)$  of the orientation $\theta_{k,0}/2\pi$ for (a) $Ra = 3.45 \times 10^{12}$ and (b) $Ra = 1.35\times 10^{14}$, for $z/L = 0.25$ (red squares), $0.50$ (green circles) and $0.75$ (blue triangles). Solid lines represent fits of Gaussian functions to the data. }
\label{fig:pdf}
\end{figure}

\begin{figure}
\centerline{\includegraphics[width=0.8 \textwidth]{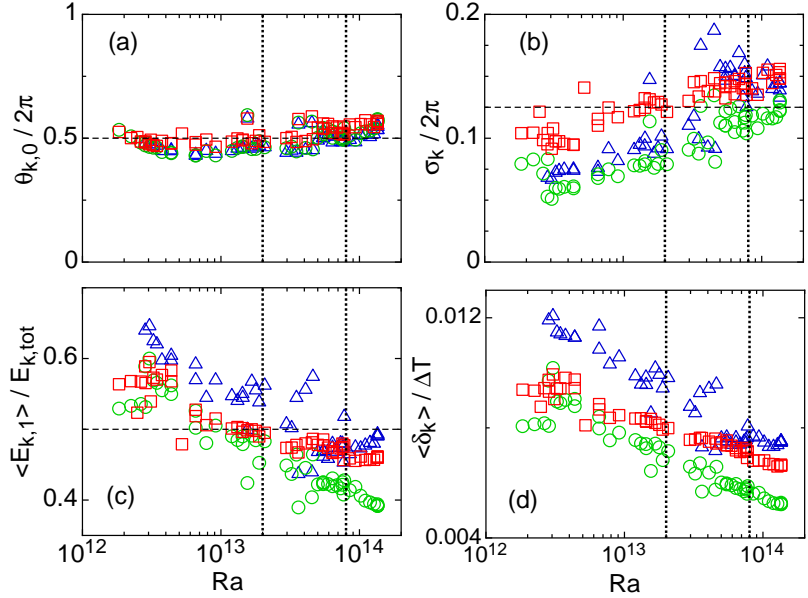}}
\caption{(a): Preferred orientation $\theta_{k,0}/(2\pi)$, (b): width $\sigma_k/(2\pi)$ of $p(\theta_k/2\pi)$, (c): LSC fundamental-mode energy $\langle E_{k,1} \rangle/E_{k,tot}$, and (d): LSC temperature amplitude $\langle \delta_k \rangle / \Delta T$ as a function of $Ra$ for $z/L = 0.25$ (red squares), $0.50$ (green circles), and $0.75$ (blue triangles). Vertical dotted lines: $Ra^*_1 = 2\times 10^{13}$ (left) and $Ra^*_2 = 8\times 10^{13}$ (right) from Fig.~\ref{fig:Re_th_Pr} below. Horizontal dashed line in (a): $\theta_{k,0} = \pi$ (West).}
\label{fig:Radept}
\end{figure}

Probability distributions $p(\theta_{k,0}/2\pi)$ for the three levels $k = b, m, t$ are shown in Fig.~\ref{fig:pdf} for (a) the classical and (b) the ultimate state. One sees a coherent orientation over the height of the sample. We fitted the function
\be
p_f(\theta_k) = p_{k,0}\exp\left[(\theta_k-\theta_{k,0})^2 / (2\sigma_k)^2 \right]
\ee
to the data for $p(\theta_k)$. The results for $\theta_{k,0}$ and $\sigma_k$ are shown in Figs.~\ref{fig:Radept}(a) and (b) as a function of $Ra$. Consistent with the predictions of \cite{BA06b} for the influence of Earth's Coriolis force, the orientation $\theta_{k,0}$ of the LSC up-flow is close to West ($\pi$) for all $Ra$. The width is slightly larger near the top and bottom of the sample than it is in the middle, but the variation is not large. The relative contribution $\langle E_{k,1}\rangle = \langle \delta_k^2 \rangle$ to the total fluctuation energy in a horizontal plane and near the side wall [Fig.~\ref{fig:Radept}(c)] and the temperature amplitude $\langle\delta_k\rangle/\Delta T$ (Fig.~\ref{fig:Radept}(d)) are also nearly independent of $z$. Although $\langle E_{1,k}\rangle/E_{tot,k} \simeq 0.5$ and $\langle\delta_k\rangle/\Delta T \simeq 0.01$ are smaller than they are for smaller $Ra$ and larger $Pr$ [see {\it e.g.} \cite{BA07_EPL,WA11c} and references therein], the coherence of all measured quantities over all three levels indicates the existence of a statistically well defined, albeit highly fluctuating, single-roll LSC.  All parameters  depend only mildly upon $Ra$. There is no strong signature of the ultimate-state transition, but the larger scatter for $Ra_1^* \alt Ra \alt Ra_2^*$ and the much smaller scatter for $Ra > Ra_2^*$ is remarkable.

\subsection{The azimuthal diffusivity of the large-scale-circulation plane}

\begin{figure}
\centerline{\includegraphics[width=0.85 \textwidth]{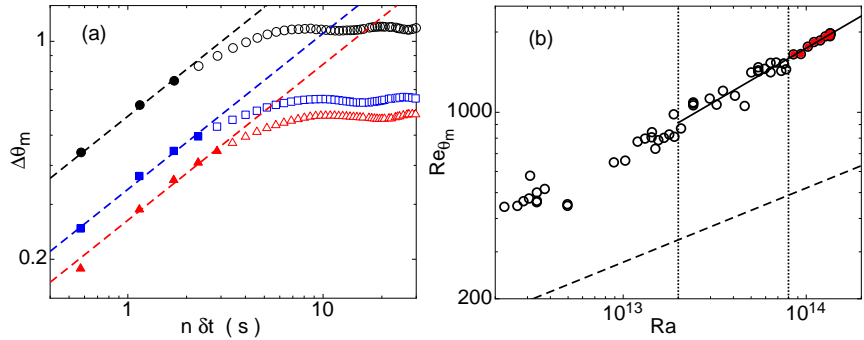}}
\caption{(a): The rms azimuthal displacement $\Delta\theta_m$ at $z/L = 0.50$ as a function of $n\delta t$ on logarithmic 
scales for $z/L = 0.5$ and $Ra =1.35 \times 10^{14}$ (black circles), $1.90 \times 10^{13}$ (blue squares) and $2.24 \times 10^{12}$ (red triangles). The dashed lines are fits of the function $\Delta\theta_m = \sqrt{D_{\theta_m} (n\delta t)}$ to the solid symbols, adjusting only $D_{\theta_m}$. 
(b): Reynolds numbers $Re_{\theta_m} \equiv L\sqrt{D_{\theta_m}/\nu}$ as a function of $Ra$ for $z/L = 0.50$. Solid line: a power-law fit (that gave an exponent of 0.40) to the circles with red dots. Dashed line: $Re_{\theta_m} = 0.0211Ra^{0.278}$ from \cite{BA06a} for $Pr = 4.38$.  Vertical dotted lines represent $Ra^*_1 = 2\times 10^{13}$ (left) and $Ra^*_2 = 8\times 10^{13}$ (right).}
\label{fig:diffusion}
\end{figure}

For a diffusive process in a system with a flat (constant) potential the root-mean-square (rms) azimuthal displacement 
\be
\Delta\theta_k \equiv \sqrt{\langle[\theta_{k,0}(t+n\delta t) - \theta_{k,0}(t)]^2\rangle}\ ,
\ee
computed over various time intervals $n\delta t$, is given by 
\be
\Delta\theta_k = \sqrt{D_{\theta_k}(n\delta t)}\ .
\label{eq:diff}
\ee
Here  $D_{\theta_k}$ is the azimuthal diffusivity, $\delta t$ is the time interval between successive data points, $n$ is a positive integer, and $\langle\rangle$ indicates a time average. 

Results for $\Delta\theta_m$ as a function of $n\delta t$ for $z/L = 0.50$ are shown in Fig.~\ref{fig:diffusion}(a) for three values of $Ra$. One sees that the square-root law holds for small $n\delta t$, but that at large time intervals $\Delta\theta$ becomes constant. This is because the azimuthal potential had a minimum rather than being flat, as can be inferred from the probability distributions $p(\theta_{m,0})$ shown in Fig.~\ref{fig:pdf}. In that case diffusion in the potential well was limited to a finite range approximately equal to $\sigma_m$  \cite[]{BA06b} [see Fig.~\ref{fig:Radept} (b)].   
One can still obtain values of $D_{\theta_m}$ from a fit of Eq.~\ref{eq:diff} to data at small $n \delta t$ such as those shown as solid symbols in Fig.~\ref{fig:diffusion}(a). The corresponding results for $Re_{\theta_m}$ are shown in Fig.~\ref{fig:diffusion}(b). 
 
\begin{figure}
\centerline{\includegraphics[width=0.95 \textwidth]{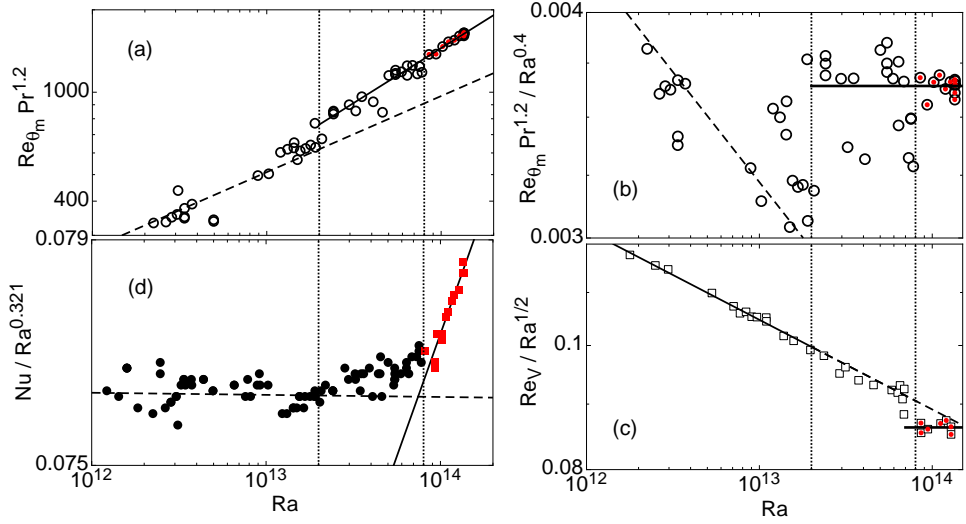}}
\caption{(a): $Re_{\theta_m}Pr^{1.2}$ as a function of $Ra$ for $z/L = 0.50$ and $Pr \simeq 0.8$. Solid line: a power-law fit with an exponent of 0.40 to the circles with red dots. Dashed line: $0.124Ra^{0.278}$ as determined using the data from \citep{BA06a} for $Pr = 4.38$. 
(b): Reduced Reynolds number $Re_{\theta_m}Pr^{1.2}/Ra^{0.4}$ as a function of Ra for $Pr \simeq 0.8$. Dashed line: an extrapolation of the results for $Pr = 4.38$ of \cite{BA06a}. The horizontal solid bar corresponds to $0.00364$. (c): The reduced velocity-fluctuation Reynolds number $Re_V/Ra^{0.5}$ as a function of $Ra$ from \cite{HGBA15b}. The solid line is the power-law fit to the data with $Ra \leq Ra^*_1 = 2 \times 10^{13}$, which gave an exponent of 0.43. The dashed line is the extrapolation of that fit beyond $Ra^*_1$. The horizontal solid bar corresponds to 0.0863. (d): The reduced Nusselt number $Nu/Ra^{0.321}$ as a function of $Ra$ on logarithmic scales. The solid line is the power-law fit $Nu = 0.0159 Ra^{0.370}$ to the red squares, and the horizontal dashed line is at $Nu/Ra^{0.321} = 0.0763$. 
In all panels vertical dotted lines represent $Ra^*_1 \simeq 2\times 10^{13}$ and $Ra^*_2 \simeq 8\times 10^{13}$.
}
\label{fig:Re_th_Pr}
\end{figure}

The dashed line in Fig.~\ref{fig:diffusion}(b) is an extrapolation of the result obtained by \cite{BA06a} for $Ra \alt 10^{11}$ and $Pr = 4.38$. Although our data are at much larger $Ra$, we note that this extrapolation can be made to pass through our data for classical RBC ($Ra \alt 2\times 10^{13}$) when $Re_{\theta_m}Pr^{1.2}$ is plotted vs. $Ra$, as in Fig.~\ref{fig:Re_th_Pr}(a). We note that the same dependence on $Pr^{1.2}$ was found also for the velocity-fluctuation Reynolds number $Re_V$ in the range $Pr \alt 7$ \cite[]{HGBA15b}.

\subsection{$Nu$, $Re_V$, and $Re_{\theta}$, and the ultimate-state transition-range from $Ra_1^*$ to $Ra_2^*$ }
\label{sec:ultimate}

In Fig.~\ref{fig:Re_th_Pr}(a) one sees that there is a change in the $Ra$ dependence of $Re_{\theta_m}Pr^{1.2}$ near $Ra = 2\times 10^{13}$ which we identify with the beginning of the transition range to the ultimate state of RBC at $Ra_1^*$ (left vertical dotted line). This becomes more obvious when the reduced data for $Re_{\theta_m}Pr^{1.2}/Ra^{0.4}$ are plotted, as in Fig.~\ref{fig:Re_th_Pr}(b). There one also sees that the scatter in the data suddenly decreases as $Ra$ exceeds $Ra^*_2 \simeq 8\times 10^{13}$, as was the case for $Re_V$ [see {\it e.g.} \cite{HFNBA12}], and for $\langle E_{k,1}\rangle/E_{k,tot}$ in Fig.~\ref{fig:Radept}(c) shown above. The origin of the scatter in the transition region is unclear.  One reason could be that  between the classical and ultimate states the  system can choose from two or more states with different boundary-layer configurations where the BLs are, spatially, neither entirely laminar nor entirely turbulent. Above the transition the fluctuations in the data are small and the state seems uniquely defined. Further  experiments are needed to better understand the transition region.

For larger $Ra$ we find $Re_{\theta_m}Pr^{1.2} \propto Ra^{0.40\pm 0.03}$, as shown by the solid black lines in Figs.~\ref{fig:Re_th_Pr}(a) and (b). A theoretical explanation of the $Ra$ and $Pr$ dependence of $Re_\theta$, both in the classical and the ultimate state, remains a challenge.

In Fig.~\ref{fig:Re_th_Pr}(c) we show $Re_V/Ra^{0.50}$ from \cite{HGBA15b}. Those data were taken in a different cell (``HPCF-IV"). While they did not give an obvious indication of $Ra^*_1$ (except for a slight increase of the scatter), they clearly show a discontinuity and a change of the $Ra$ dependence at $Ra^*_2 \simeq 8\times 10^{13}$. The $Ra$ dependence $Re_V \propto Re^{0.50 \pm 0.02}$ is in excellent agreement with the prediction by \cite{GL11} for the mean-flow Reynolds number $Re_U$ in the ultimate state.

Figure~\ref{fig:Re_th_Pr}(d) shows new $Nu$ measurements for the same sample used here (``HPCF-IVb") for the diffusivity measurements. One sees that $Nu/Ra^{0.321}$ is constant in the classical state, and begins to gently increase for $Ra > Ra_1^*$. For $Ra > Ra_2^*$ (red squares) we find $Nu \propto Ra^{0.37 \pm 0.02}$, also consistent with the prediction  for the ultimate state \cite[]{GL11}.

The data in Fig.~\ref{fig:Re_th_Pr} clearly identify a transition to a new state of the system, with a transition range starting at $Ra_1^* \simeq 2\times 10^{13}$. It is important to decide whether this $Ra$ value is consistent with the BL shear instability associated with the ultimate-state transition. Since both the LSC and fluctuations are expected to contribute to the BL instability, we use the measured $Re_{eff} \equiv \sqrt{Re_U^2 + Re_V^2} = 1.00\times Ra^{0.436}$ \cite[]{HGBA15b} for $\Gamma = 1.00$ which, for $Ra_1^* \simeq 2\times 10^{13}$, gives $Re_{eff} = 6.3\times10^5$ and thus yields the shear Reynolds number \cite[]{GL02} $Re_s^* \simeq  0.48 \sqrt{Re_{eff}} \simeq 380$. While it is difficult to estimate the overall uncertainty of this value, it is generally consistent with the expected $Re_s^* \simeq 400$ \cite[]{LL87}.

\section{Non-Oberbeck-Boussinesq effects and the Skrbek-Urban claim}
\label{sec:NOB}

Real systems will deviate from the OB approximation. Consequently, $T_c$ may differ from $T_m$ and various global and local properties may be affected.
While in most experiments these effects are small and often negligible, \cite{SU15} (SU) claimed that  they have a {\it large} influence on the G\"ottingen $Nu$ measurements, and that the observation of the ultimate-state transition based on those measurements may be illusory and a misinterpretation of non-OB effects  (they make similar claims regarding the Grenoble measurements). In this section we shall show that the 
arguments of SU are flawed and that there is indeed no evidence that non-OB effects alter the original conclusions regarding the ultimate-state transition as discussed in part in $\S$ \ref{sec:ultimate}.

To lowest order non-OB effects have no influence on $Nu$ when the fluid properties are evaluated at $T_m$ \cite[]{ABFFGL06}. The physics of this is simple. In the classical state one may, to a good approximation, regard $Nu$ as proportional to the inverse of the sum of two resistors in series and carrying the same heat current. One ($R_b$) corresponds to the bottom and the other ($R_t$) to the top BL, and the turbulent bulk between the BLs is taken to be a perfect conductor. While one resistance (that of the thicker BL) is increased slightly by the non-OB conditions, the other is decreased slightly by a nearly equal amount and their sum, and thus $Nu$, remains nearly the same. This is shown theoretically in $\S$ 4 of  \cite{ABFFGL06}. It is not only a theoretical result, but also was demonstrated experimentally by the same authors. Although one might think that this argument does not apply to the ultimate state where the turbulent BLs extend deep into the interior \cite[]{GL11,GL12}, in practice also there one finds most of the temperature drop across thin layers near the plates \cite[]{ABFGHLSV12} and the ``two resistor model" remains reasonable at least for $Ra$ not too much larger than $Ra_2^*$.   

For the determination of $Nu$ from the G\"ottingen measurements \cite[]{HFNBA12,AHFB12,HFBA12,HGBA15b}, which revealed the transition to the ultimate state, careful experimental consideration had been given to the need to correct for non-OB effects. Possible differences due to the definition of $Ra$  were quite small since $T_c$ was very close to $T_m$ [\cite{HFBA12} figure 1(b); see also \cite{HFNBA13}], with $(T_c - T_m)/T_m \alt 0.002$ (temperatures are in Kelvin). With such a small change of the absolute temperature the properties of a gas do not vary significantly, and it does not matter for instance whether $Ra$ is evaluated using the properties at $T_m$ as is customary and preferred theoretically or at $T_c$ as suggested by SU. Further, it had been shown experimentally that non-OB effects did not influence the measurements significantly by comparing data at the same $Ra$ taken with relatively large $\Delta T$ where non-OB effects are larger (by using relatively low pressures) with others taken with smaller $\Delta T$ where non-OB effects are smaller (measured at  larger pressures) [figure B.1(b) of \cite{AHFB12}].
   
Contrary to the above, SU claim that  non-OB effects have a {\it large} influence on the G\"ottingen $Nu$ data. They state that \\ 
"the claims of observing transition to the ultimate state of Rayleigh-B\'enard convection
could indeed be related to non-Oberbeck-Boussinesq effects". \\
They then propose a different method of analysis (the ``SU model") which, they argue, corrects for these large non-OB effects. Below we show that the arguments by SU have two flaws.

The first flaw is that they arbitrarily decided that it would be better to evaluate the fluid properties at  $T_c$ (either experimentally measured or derived from a crude model) rather than at $T_m$. It has been known since the trailblazing work of Fritz Busse \cite[]{Bu67} [see also \cite{ADOP09}]  for convection near onset and was extended to the turbulent state by \cite{ABFFGL06} that this is inappropriate. A more systematic approach is to retain $T_m$ as the only {\it a priori} known reference temperature, and to treat non-OB effects with this choice as perturbations of the OB state. This systematic procedure will then yield a value of $T_c - T_m$ {\it a posteriori}. However, as indicated above and shown also by \cite{HFNBA13}, choosing $T_c$ instead has in practice virtually no consequences for the G\"ottingen $Nu$ and $Ra$ values because $|(T_c - T_m)|/T_m$ was small. Nonetheless, there is no foundation for SU'€™s choice of $T_c$ rather than $T_m$ and it goes contrary to a large body of well established literature.

The second flaw has a greater impact on the $Nu$ and $Ra$ values. SU claim that in the G\"ottingen experiments\\ 
``$T_t$ values are dangerously close to the SVP line (representing the equilibrium first-order liquid-€"gas phase transition where due to fluctuations condensation/evaporation could take place)".\\ 
Please let us re-state the nature of first-order phase transitions. There is no anomalous increase of fluctuation intensities, and all properties vary only slowly with temperature until the very point where the transition temperature is crossed and latent-heat contributions to $Nu$ occur. Thus the claim by SU is unfounded. SU then move on to claim\\
``We can further strengthen our argument by pointing out the work of Zhong, Funfschilling \& Ahlers (2009), showing that the heat transfer efficiency would be considerably enhanced if condensation/evaporation processes were to take place in the vicinity of the SVP line at the top plate of the SF6 cell."\\
This is a misinterpretation of the mentioned paper. \cite{ZFA09} showed that, with decreasing  mean temperature,  there is NO excess contribution to the heat transport until the very point at which the top temperature $T_t$ crosses the vapor-pressure curve at $T_\phi(P)$. When $T_\phi$ is crossed by $T_t$ (which never happened in the G\"ottingen measurements), a continuous but sharp transition takes place to a state where the heat transport increases linearly with $T_\phi - T_t$.

SU then continue to state\\
``Upon increasing Ra, the top half of the cell becomes affected by the non-Oberbeck-Boussinesq effects, while the bottom half, if lying sufficiently far away from the SVP curve, does not."\\
Needless to say, this is incorrect. All physical systems are influenced by non-OB effects; the question  only is whether these effects influence {\it significantly} a given measured quantity. While the non-OB effect is indeed stronger at the top of the cell, it does not vanish in the bottom and for the G\"ottingen data (by any reasonable measure) is smaller by less than a factor of two or so.

SU then argue that\\
``one can avoid the non-Oberbeck-Boussinesq effects in the top half of the cell, by replacing it
with the inverted (with respect to $T_c$) nearly Oberbeck-Boussinesq bottom half".\\
This argument is at odds with the beautiful cancellation between the increase of $R_b$ and decrease of $R_t$ characteristic of non-OB systems as discussed above, and creates a qualitatively new fictitious system in which Nu is proportional to $1/(2 R_b)$ rather than to $1/(R_b + R_t)$ as is the case in the physical system! To make things worse, they then replace $\Delta T \equiv T_b - T_t$  by $\Delta T_{eff} = 2(T_b - T_c)$. This SU model yields values of $Nu$ and $Ra$ which are unrelated to reality. 

It is amusing (but unrelated to the present discussion) that a non-OB system with BLs that have properties which are reflection symmetric about the horizontal mid plane of the sample does indeed exist. It occurs above the critical pressure of a fluid when, at constant pressure, the top and bottom temperatures are on opposite sides of and about equidistant from the temperature on the critical isochore \cite[]{ACFFGLS08,ADOP09,BSS10,BS12}.

Beyond their claims for the influence of non-OB effects on $Nu$, SU also state that\\
``It is important to emphasize that not only the Nusselt number scaling, but the scaling of any other independently measured quantity, such as the Reynolds number $Re = Re(Ra)$, that might have displayed `phase transitions` spuriously interpreted as independent confirmation of the transition to the ultimate regime, will change, as well."\\
We do not agree with this statement. $Nu$ is determined largely by the boundary layers with their steep thermal gradients, while recent $Re_V$ measurements \cite[]{HGBA15b}, as well as the diffusivity measurements presented here, are properties of the bulk which depend only on the local temperature which is nearly uniform and close to $T_c$ (and $T_c$ in turn is nearly equal to $T_m$). In the SU model the major influence on $Re(Ra)$ arises from using $\Delta T_{eff}$ rather than $\Delta T$ to calculate $Ra$ and, as just stated, this use leads to an unphysical model for RBC.

From the above discussion one can see that the data manipulation of  SU 
leads to a fictitious system which is unrelated to those investigated in Grenoble or G\"ottingen. 
Thus we see no validity in questioning the claim, based on the G\"ottingen data, that a transition to a new state of RBC has been observed and that this state has been at least partially characterized.

\section{Summary and Conclusions}

In this paper we presented new measurements of the temperature amplitude and of the azimuthal diffusivity of the large-scale circulation for a cylindrical sample with $\Gamma = 1.00$ which show the ultimate-state transition-range from $Ra^*_1 \simeq 2\times 10^{13}$ to $Ra_2^* \simeq 8\times 10^{13}$, in agreement with findings based on measurements of $Nu$ and of the velocity-fluctuation Reynolds number $Re_V$.

We also showed that the claim by \cite{SU15} that ``observing transition to the ultimate state could indeed be related to non-Oberbeck-Boussinesq effects" is without any foundation. 

Finally we note that turbulent circular Couette flow (CCF) was shown to have equations of motion equivalent to those of turbulent RBC \cite[]{EGL07b}. Experiments for this system yielded a ``Nusselt number" $Nu_\omega \propto Ta^{0.38}$  which characterizes the angular velocity transport \cite[]{GHBSL11}. Here the Taylor number $Ta$ plays a role equivalent to $Ra$ in RBC. A so-called ``wind Reynolds number", analogous to the Reynolds number in RBC,  was found to be proportional to $Ta^{0.50}$  \cite[]{HGGSL12}. Both of these results agree with the analogous ultimate-state results for RBC shown in Figs.~\ref{fig:Re_th_Pr}(c) and (d). Thus there is strong evidence that both RBC and CCF, at high enough driving, enter equivalent states which have properties expected of an ultimate state with turbulent boundary layers \cite[]{GL11}. Further it is worth noting that for both systems the ultimate state is entered at about the same value, close to 400, of the boundary-layer shear Reynolds-number [for RBC see $\S$ \ref{sec:ultimate} in the present paper and \cite{HFNBA12}; for CCF see  $\S$ 5 and Fig. 14(b) of \cite{OSGVL13}, as well as \cite{GHGSL12,OPVGL14,GLS16}] where the transition to a turbulent BL is expected \cite[]{LL87}.

\section{Acknowledgments}
\begin{acknowledgments}
We are grateful to the Max-Planck-Society and the Volkswagen Stiftung, whose support made the establishment of the Uboot facility and the experiments possible. We thank the Deutsche Forschungsgemeinschaft (DFG) for financial support through SFB963: ``Astrophysical Flow Instabilities and Turbulence".  X.H. acknowledges support through the Chinese Thousand Young Talents Program. The work of G.A. was supported in part by the U.S National Science Foundation through Grant DMR11-58514.
\end{acknowledgments}

\bibliographystyle{jfm}

%\bibliography{./refs}

\begin{thebibliography}{49}
\expandafter\ifx\csname natexlab\endcsname\relax\def\natexlab#1{#1}\fi

\bibitem[Ahlers(2009)]{Ah09}
{\sc Ahlers, G.} 2009 Turbulent convection. {\em Physics\/} {\bf 2}, 74--1--7.

\bibitem[Ahlers {\em et~al.\/}(2012{\natexlab{{\em a\/}}})Ahlers, Bodenschatz,
  Funfschilling, Grossmann, He, Lohse, Stevens \& Verzicco]{ABFGHLSV12}
{\sc Ahlers, G., Bodenschatz, E., Funfschilling, D., Grossmann, S., He, X.,
  Lohse, D., Stevens, R. \& Verzicco, R.} 2012{\natexlab{{\em a\/}}}
  Logarithmic temperature profiles in turbulent {Rayleigh-B\'enard} convection.
  {\em Phys. Rev. Lett.\/} {\bf 109}, 114501--1--5.

\bibitem[Ahlers {\em et~al.\/}(2014)Ahlers, Bodenschatz \& He]{ABH14}
{\sc Ahlers, G., Bodenschatz, E. \& He, X.} 2014 Logarithmic temperature
  profiles of turbulent {Rayleigh-B\'enard} convection in the classical and
  ultimate state for a {P}randtl number of 0.8. {\em J. Fluid Mech.\/} {\bf
  758}, 436--467.

\bibitem[Ahlers {\em et~al.\/}(2006)Ahlers, Brown, {{Fontenele Araujo}},
  Funfschilling, Grossmann \& Lohse]{ABFFGL06}
{\sc Ahlers, G., Brown, E., {{Fontenele Araujo}}, F., Funfschilling, D.,
  Grossmann, S. \& Lohse, D.} 2006 {{Non-Oberbeck-Boussinesq}} effects in
  strongly turbulent {{Rayleigh-B\'enard}} convection. {\em J. Fluid Mech.\/}
  {\bf 569}, 409--445.

\bibitem[Ahlers {\em et~al.\/}(2008)Ahlers, Calzavarini, {Fontenele Araujo},
  Funfschilling, Grossmann, Lohse \& Sugiyama]{ACFFGLS08}
{\sc Ahlers, G., Calzavarini, E., {Fontenele Araujo}, F., Funfschilling, D.,
  Grossmann, S., Lohse, D. \& Sugiyama, K.} 2008 Non-{Oberbeck}-{Boussinesq}
  effects in turbulent thermal convection in ethane close to the critical
  point. {\em Phys. Rev. E\/} {\bf 77}, 046302--1--16.

\bibitem[Ahlers {\em et~al.\/}(2009{\natexlab{{\em a\/}}})Ahlers, Dressel, Oh
  \& Pesch]{ADOP09}
{\sc Ahlers, G., Dressel, B., Oh, J. \& Pesch, W.} 2009{\natexlab{{\em a\/}}}
  Strong non-{Boussinesq} effects near the onset of convection in a fluid near
  its critical point. {\em J. Fluid Mech.\/} {\bf 642}, 15--48.

\bibitem[Ahlers {\em et~al.\/}(2009{\natexlab{{\em b\/}}})Ahlers, Funfschilling
  \& Bodenschatz]{AFB09}
{\sc Ahlers, G., Funfschilling, D. \& Bodenschatz, E.} 2009{\natexlab{{\em
  b\/}}} Transitions in heat transport by turbulent convection for {$Pr = 0.8$}
  and {$10^{11} \leq Ra \leq 10^{15}$}. {\em New J. Phys.\/} {\bf 11},
  123001--1--18.

\bibitem[Ahlers {\em et~al.\/}(2009{\natexlab{{\em c\/}}})Ahlers, Grossmann \&
  Lohse]{AGL09}
{\sc Ahlers, G., Grossmann, S. \& Lohse, D.} 2009{\natexlab{{\em c\/}}} Heat
  transfer and large scale dynamics in turbulent {Rayleigh}-{B\'enard}
  convection. {\em Rev. Mod. Phys.\/} {\bf 81}, 503--538.

\bibitem[Ahlers {\em et~al.\/}(2012{\natexlab{{\em b\/}}})Ahlers, He,
  Funfschilling \& Bodenschatz]{AHFB12}
{\sc Ahlers, G., He, X., Funfschilling, D. \& Bodenschatz, E.}
  2012{\natexlab{{\em b\/}}} Heat transport by turbulent {Rayleigh-B\'enard}
  convection for {$Pr \simeq 0.8$} and {$3\times 10^{12} \alt Ra\ \alt
  10^{15}$}: Aspect ratio $\gamma = 0.50$. {\em New J. Phys.\/} {\bf 14},
  103012--1--39.

\bibitem[Boussinesq(1903)]{Bo03}
{\sc Boussinesq, J.} 1903 {\em Theorie analytique de la chaleur, Vol. 2\/}.
  Paris: Gauthier-Villars.

\bibitem[Brown \& Ahlers(2006{\natexlab{{\em a\/}}})]{BA06b}
{\sc Brown, E. \& Ahlers, G.} 2006{\natexlab{{\em a\/}}} Effect of the
  {{Earth's}} {{Coriolis}} force on turbulent {{Rayleigh-B{\'e}nard}}
  convection in the laboratory. {\em Phys. Fluids\/} {\bf 18}, 125108--1--15.

\bibitem[Brown \& Ahlers(2006{\natexlab{{\em b\/}}})]{BA06a}
{\sc Brown, E. \& Ahlers, G.} 2006{\natexlab{{\em b\/}}} Rotations and
  cessations of the large-scale circulation in turbulent
  {{Rayleigh-B{\'e}nard}} convection. {\em J. Fluid Mech.\/} {\bf 568},
  351--386.

\bibitem[Brown \& Ahlers(2007{\natexlab{{\em a\/}}})]{BA07a}
{\sc Brown, E. \& Ahlers, G.} 2007{\natexlab{{\em a\/}}} Large-scale
  circulation model of turbulent {{Rayleigh-B{\'e}nard}} convection. {\em Phys.
  Rev. Lett.\/} {\bf 98}, 134501--1--4.

\bibitem[Brown \& Ahlers(2007{\natexlab{{\em b\/}}})]{BA07_EPL}
{\sc Brown, E. \& Ahlers, G.} 2007{\natexlab{{\em b\/}}} Temperature gradients,
  and search for non-{Boussinesq} effects, in the interior of turbulent
  {Rayleigh}-{B\'enard} convection. {\em Europhys. Lett.\/} {\bf 80},
  14001--1--6.

\bibitem[Brown \& Ahlers(2008)]{BA08a}
{\sc Brown, E. \& Ahlers, G.} 2008 A model of diffusion in a potential well for
  the dynamics of the large-scale circulation in turbulent
  {Rayleigh}-{B{\'e}nard} convection. {\em Phys. Fluids\/} {\bf 20},
  075101--1--16.

\bibitem[Brown {\em et~al.\/}(2005)Brown, Nikolaenko \& Ahlers]{BNA05}
{\sc Brown, E., Nikolaenko, A. \& Ahlers, G.} 2005 Reorientation of the
  large-scale circulation in turbulent {{Rayleigh-B{\'e}nard}} convection. {\em
  Phys. Rev. Lett.\/} {\bf 95}, 084503--1--4.

\bibitem[Burnishev {\em et~al.\/}(2010)Burnishev, Segre \& Steinberg]{BSS10}
{\sc Burnishev, Y., Segre, E. \& Steinberg, V.} 2010 Strong symmetrical
  non-{Oberbeck-Boussinesq} turbulent convection and the role of
  compressibility. {\em Phys. Fluids\/} {\bf 22}, 035108--1--15.

\bibitem[Burnishev \& Steinberg(2012)]{BS12}
{\sc Burnishev, Y. \& Steinberg, V.} 2012 Statistics and scaling properties of
  temperature field in symmetrical non-{Oberbeck-Boussinesq} turbulent
  convection. {\em Phys. Fluids\/} {\bf 24}, 045102--1--21.

\bibitem[Busse(1967)]{Bu67}
{\sc Busse, F.~H.} 1967 The stability of finite amplitude cellular convection
  and its relation to an extremum principle. {\em J. Fluid Mech.\/} {\bf 30},
  625--649.

\bibitem[Chavanne {\em et~al.\/}(1997)Chavanne, Chilla, Castaing, Hebral,
  Chabaud \& Chaussy]{CCCHCC97}
{\sc Chavanne, X., Chilla, F., Castaing, B., Hebral, B., Chabaud, B. \&
  Chaussy, J.} 1997 Observation of the ultimate regime in {{Rayleigh-B\'enard}}
  convection. {\em Phys. Rev. Lett.\/} {\bf 79}, 3648--3651.

\bibitem[Chavanne {\em et~al.\/}(2001)Chavanne, Chilla, Chabaud, Castaing \&
  Hebral]{CCCCH01}
{\sc Chavanne, X., Chilla, F., Chabaud, B., Castaing, B. \& Hebral, B.} 2001
  Turbulent {{Rayleigh-B\'enard}} convection in gaseous and liquid he. {\em
  Phys. Fluids\/} {\bf 13}, 1300--1320.

\bibitem[Chill\`a \& Schumacher(2012)]{CS12}
{\sc Chill\`a, F. \& Schumacher, J.} 2012 New perspectives in turbulent
  {Rayleigh-B\'enard} convection. {\em Eur. Phys. J. E\/} {\bf 35}, 58--1--25.

\bibitem[Eckhardt {\em et~al.\/}(2007)Eckhardt, Grossmann \& Lohse]{EGL07b}
{\sc Eckhardt, B., Grossmann, S. \& Lohse, D.} 2007 Torque scaling in turbulent
  {Taylor-Couette} flow between independently rotating cylinders. {\em J. Fluid
  Mech.\/} {\bf 581}, 221--250.

\bibitem[van Gils {\em et~al.\/}(2012)van Gils, Huisman, Grossmann, Sun \&
  Lohse]{GHGSL12}
{\sc van Gils, D., Huisman, S., Grossmann, S., Sun, C. \& Lohse, D.} 2012
  Optimal {Taylor-Couette} turbulence. {\em J. Fluid Mech.\/} {\bf 706},
  118--149.

\bibitem[van Gils {\em et~al.\/}(2011)van Gils, Huisman, Briggert, Sun \&
  Lohse]{GHBSL11}
{\sc van Gils, D. P.~M., Huisman, S.~G., Briggert, G.-W., Sun, C. \& Lohse, D.}
  2011 Torque scaling in turbulent taylor-couette flow with co- and
  counterrotating cylinders. {\em Phys. Rev. Lett.\/} {\bf 106}, 024502--1--4.

\bibitem[Grossmann \& Lohse(2002)]{GL02}
{\sc Grossmann, S. \& Lohse, D.} 2002 Prandtl and {{Rayleigh}} number
  dependence of the {{Reynolds}} number in turbulent thermal convection. {\em
  Phys. Rev. E\/} {\bf 66}, 016305--1--6.

\bibitem[Grossmann \& Lohse(2011)]{GL11}
{\sc Grossmann, S. \& Lohse, D.} 2011 Multiple scaling in the ultimate regime
  of thermal convection. {\em Phys. Fluids\/} {\bf 23}, 045108--1--6.

\bibitem[Grossmann \& Lohse(2012)]{GL12}
{\sc Grossmann, S. \& Lohse, D.} 2012 Logarithmic temperature profiles in the
  ultimate regime of thermal convection. {\em Phys. Fluids\/} {\bf 24},
  125103--1--8.

\bibitem[Grossmann {\em et~al.\/}(2016)Grossmann, Lohse \& Sun]{GLS16}
{\sc Grossmann, S., Lohse, D. \& Sun, C.} 2016 High {Reynolds} number
  {Taylor-Couette} turbulence. {\em Annu. Rev. Fluid Mech.\/} {\bf 48}, 53--80.

\bibitem[He {\em et~al.\/}(2012{\natexlab{{\em a\/}}})He, Funfschilling,
  Bodenschatz \& Ahlers]{HFBA12}
{\sc He, X., Funfschilling, D., Bodenschatz, E. \& Ahlers, G.}
  2012{\natexlab{{\em a\/}}} Heat transport by turbulent {Rayleigh-B\'enard}
  convection for ${Pr}\simeq 0.8$ and $4\times 10^{11} \alt {Ra} \alt
  2\times10^{14}$: Ultimate-state transition for aspect ratio $\gamma = 1.00$.
  {\em New J. Phys.\/} {\bf 14}, 063030--1--15.

\bibitem[He {\em et~al.\/}(2012{\natexlab{{\em b\/}}})He, Funfschilling,
  Nobach, Bodenschatz \& Ahlers]{HFNBA12}
{\sc He, X., Funfschilling, D., Nobach, H., Bodenschatz, E. \& Ahlers, G.}
  2012{\natexlab{{\em b\/}}} Transition to the ultimate state of turbulent
  {Rayleigh-B\'enard} convection. {\em Phys. Rev. Lett.\/} {\bf 108},
  024502--1--5.

\bibitem[He {\em et~al.\/}(2013)He, Funfschilling, Nobach, Bodenschatz \&
  Ahlers]{HFNBA13}
{\sc He, X., Funfschilling, D., Nobach, H., Bodenschatz, E. \& Ahlers, G.} 2013
  Comment on ``{Effect} of boundary layers asymmetry on heat transfer
  efficiency in turbulent {Rayleigh- B\'enard} convection at very high
  {Rayleigh} numbers by {Urban} {\it et al.}?. {\em Phys. Rev. Lett.\/} {\bf
  110}, 199401--1--1.

\bibitem[He {\em et~al.\/}(2014)He, {van Gils}, Bodenschatz \& Ahlers]{HGBA13}
{\sc He, X., {van Gils}, D., Bodenschatz, E. \& Ahlers, G.} 2014 Logarithmic
  spatial variations and universal $f^{-1}$ power spectra of temperature
  fluctuations in turbulent {Rayleigh-B\'enard} convection. {\em Phys. Rev.
  Lett.\/} {\bf 112}, 174501--1--5.

\bibitem[He {\em et~al.\/}(2015)He, {van Gils}, Bodenschatz \& Ahlers]{HGBA15b}
{\sc He, X., {van Gils}, D., Bodenschatz, E. \& Ahlers, G.} 2015 Reynolds
  numbers and the elliptic approximation near the ultimate state of turbulent
  {Rayleigh-B\'enard} convection. {\em New J. Phys.\/} {\bf 17}, 063028--1--26.

\bibitem[Huisman {\em et~al.\/}(2012)Huisman, van Gils, Grossmann, Sun \&
  Lohse]{HGGSL12}
{\sc Huisman, S.~G., van Gils, D. P.~M., Grossmann, S., Sun, C. \& Lohse, D.}
  2012 Ultimate turbulent {Taylor-Couette} flow. {\em Phys. Rev. Lett.\/} {\bf
  108}, 024501--1--4.

\bibitem[Kadanoff(2001)]{Ka01}
{\sc Kadanoff, L.~P.} 2001 Turbulent heat flow: Structures and scaling. {\em
  Phys. Today\/} {\bf 54}~(8), 34--39.

\bibitem[Kraichnan(1962)]{Kr62}
{\sc Kraichnan, R.~H.} 1962 Turbulent thermal convection at arbritrary
  {{Prandtl}} number. {\em Phys. Fluids\/} {\bf 5}, 1374--1389.

\bibitem[Landau \& Lifshitz(1987)]{LL87}
{\sc Landau, L.~D. \& Lifshitz, E.~M.} 1987 {\em Fluid Mechanics\/}. Oxford:
  Pergamon Press.

\bibitem[Lohse \& Xia(2010)]{LX10}
{\sc Lohse, D. \& Xia, K.-Q.} 2010 Small-scale properties of turbulent
  {Rayleigh-B{\'e}nard} convection. {\em Annu. Rev. Fluid Mech.\/} {\bf 42},
  335--364.

\bibitem[Oberbeck(1879)]{Ob79}
{\sc Oberbeck, A.} 1879 \"Uber die {{W\"armeleitung}} der {{Fl\"ussigkeiten}}
  bei {{Ber\"ucksichtigung}} der {{Str\"omungen}} infolge von
  {{Temperaturdifferenzen}}. {\em Ann. Phys. Chem.\/} {\bf 7}, 271--292.

\bibitem[Ostilla {\em et~al.\/}(2013)Ostilla, Stevens, Grossmann, Verzicco \&
  Lohse]{OSGVL13}
{\sc Ostilla, R., Stevens, R., Grossmann, S., Verzicco, R. \& Lohse, D.} 2013
  Optimal {Taylor-Couette} flow: direct numerical simulation. {\em J. Fluid
  Mech.\/} {\bf 719}, 14--46.

\bibitem[Ostilla-Monico {\em et~al.\/}(2014)Ostilla-Monico, {van der Poel},
  Verzicco, Grossmann \& Lohse]{OPVGL14}
{\sc Ostilla-Monico, R., {van der Poel}, E., Verzicco, R., Grossmann, S. \&
  Lohse, D.} 2014 Exploring the phase diagram of fully turbulent
  {Taylor-Couette} flow. {\em J. Fluid Mech.\/} {\bf 761}, 1--26.

\bibitem[Roche {\em et~al.\/}(2010)Roche, Gauthier, Kaiser \& Salort]{RGKS10}
{\sc Roche, P.-E., Gauthier, F., Kaiser, R. \& Salort, J.} 2010 On the
  triggering of the ultimate regime of convection. {\em New J. Phys.\/} {\bf
  12}, 085014--1--26.

\bibitem[Skrbek \& Urban(2015)]{SU15}
{\sc Skrbek, L. \& Urban, P.} 2015 Has the ultimate state of turbulent thermal
  convection been observed? {\em J. Fluid Mech.\/} {\bf 785}, 270--282.

\bibitem[Spiegel(1971)]{Sp71}
{\sc Spiegel, E.~A.} 1971 Convection in stars. {\em Ann. Rev. Astron.
  Astrophys.\/} {\bf 9}, 323--352.

\bibitem[Stevens {\em et~al.\/}(2011)Stevens, Clercx \& Lohse]{SCL11}
{\sc Stevens, R.~J., Clercx, H.~J. \& Lohse, D.} 2011 Effect of plumes on
  measuring the large scale circulation. {\em Phys. Fluids\/} {\bf 23},
  095110--1--11.

\bibitem[Wei \& Ahlers(2014)]{WA14}
{\sc Wei, P. \& Ahlers, G.} 2014 Logarithmic temperature distribution in the
  bulk of turbulent {Rayleigh-B\'enard} convection for a {Prandtl} number of
  12.3. {\em J. Fluid Mech.\/} {\bf 758}, 809--830.

\bibitem[Weiss \& Ahlers(2011)]{WA11c}
{\sc Weiss, S. \& Ahlers, G.} 2011 The large-scale flow structure in turbulent
  rotating {Rayleigh-B\'enard} convection. {\em J. Fluid Mech.\/} {\bf 688},
  461--492.

\bibitem[Zhong {\em et~al.\/}(2009)Zhong, Funfschilling \& Ahlers]{ZFA09}
{\sc Zhong, J.-Q., Funfschilling, D. \& Ahlers, G.} 2009 Enhanced heat
  transport by turbulent two-phase {Rayleigh-B\'enard} convection. {\em Phys.
  Rev. Lett.\/} {\bf 102}, 124501--1--4.

\end{thebibliography}

\end{document}